\documentclass[aps,prb,amsmath,amssymb,twocolumn,10pt,superscriptaddress]{revtex4-2}

\usepackage{graphicx}
\usepackage{dcolumn}
\usepackage{bm}
\usepackage{times}
\usepackage[colorlinks,citecolor=blue]{hyperref}
\usepackage{xcolor}
\usepackage{kotex}

\usepackage[normalem]{ulem}\graphicspath{{figures/}}

\newcommand\redout{\bgroup\markoverwith{\textcolor{red}{\rule[.5ex]{2pt}{0.4pt}}}\ULon}

\begin{document}

\preprint{APS/123-QED}

\title{Dynamics of one-dimensional spin models via complex-time evolution of tensor networks}
\author{Jeong Hyeok Cha}
\affiliation{Department of Semiconductor Physics, Kangwon National University, Chuncheon 24341, Republic of Korea}
\affiliation{Interdisciplinary Program in Earth Environmental System Science and Engineering, Kangwon National University, Chuncheon 24341, Republic of Korea}

\author{Hyun-Yong Lee}
\email{hyunyong@korea.ac.kr}
\affiliation{Division of Semiconductor Physics, Korea University, Sejong 30019, Korea}
\affiliation{Department of Applied Physics, Graduate School, Korea University, Sejong 30019, Korea}

\author{Heung-Sik Kim}%
\email{heungsikim@kangwon.ac.kr}
\affiliation{Department of Semiconductor Physics, Kangwon National University, Chuncheon 24341, Republic of Korea}
\affiliation{Institute of Quantum Convergence Technology, Kangwon National University, Chuncheon 24341, Republic of Korea}
\affiliation{Interdisciplinary Program in Earth Environmental System Science and Engineering, Kangwon National University, Chuncheon 24341, Republic of Korea}

\date{\today}

\begin{abstract}
Studying the real-time dynamics of strongly correlated systems poses significant challenges, which have recently become more manageable thanks to advances in density matrix renormalization group (DMRG) and tensor network methods. A notable development in this area is the introduction of a complex-time evolution scheme for tensor network states, originally suggested for solving Anderson impurity model and designed to suppress the growth of entanglement under time evolution. In this study, we employ the complex-time evolution scheme to investigate the dynamics of one-dimensional spin systems, specifically the transverse-field Ising model (TFIM) and the XXZ model. Our analysis revisits the dynamic critical exponent $z$ of the TFIM and explores the dynamical structure factor in both gapped and gapless states of the XXZ model. Importantly, the complex-time evolution reproduces the results of real-time evolution while mitigating the rapid growth of quantum entanglement typically associated with the latter. These results demonstrate that the combination of complex-time evolution and extrapolation provides a robust and efficient framework for studying the dynamics of complex quantum systems, enabling more comprehensive insights into their behavior.
\end{abstract}

\maketitle

\section{\label{sec:level1}Introduction}
Density matrix renormalization group (DMRG) 
\cite{White1992}
and its extension to the tensor network method
\cite{Ostulund1995,Dukelsky1998,Vidal2003,Vidal2004,Verstraete2004} have enabled numerically accurate descriptions of low-dimensional systems. In addition to obtaining ground-state wavefunctions, real-time evolutions via the time-evolving block decimation (TEBD) or time-dependent variational principle (TDVP) can be applied\cite{White2004,Haegeman2011}, enabling extensive research into the dynamics of numerous low-dimensional many-body systems\cite{dynamics_citation1,dynamics_citation2,dynamics_citation3,dynamics_citation4,TNreview_annrev}.

One bottleneck in studying the real-time dynamics of many-body systems is the rapid increase in entanglement during time evolution\cite{calabrese2004,Schuch2008}. As a result, it hinders the study of low-frequency spectral dynamics.
To address this issue several methods have been proposed, such as linear prediction method and recursion approach, to name a few \cite{Barthel2009,White2008,Zalatel2015,Tian2021}. 
While Green's function and the spectral function in the frequency domain can be computed via dynamical DMRG methods \cite{Jeckelmann2002}, time-evolution formalism can be beneficial for the study of nonequilibrium dynamics \cite{Wolf2014}.

Recently, a scheme to perform time evolution along complex-time contour, from which real-time dynamics can be restored, has been suggested to suppress the increase in entanglement during the time evolution process and to provide an efficient way to handle dynamics of computationally demanding systems\cite{cao2311dynamical,Grundner2024}. 
This can be applied to obtain low-frequency real-time spectra of the Anderson impurity model (AIM), which can be applied to realize cost-effective dynamical mean-field theory calculations directly on the real-frequency axis without any analytic continuation \cite{cao2311dynamical,Grundner2024}. While the original work focused on employing their complex-time evolution scheme in solving the AIM for electronic structure applications, this approach can be effectively applied to a broader range of calculations, such as quench dynamics of spin systems in the vicinity of quantum critical points where the entanglement growth is often overwhelmingly fast \cite{Czischek2018}. 
We investigate the dynamics of the spin-1/2 one-dimensional transverse field Ising model (TFIM) and the XXZ model under external fields, and demonstrate that complex-time evolution can simulate real-time dynamics with significantly lower computational cost while maintaining reasonable accuracy.

Firstly, we apply the complex-time evolution and finite-size scaling to the TFIM to extract the dynamical critical exponent $z$ at the critical point. Conventionally, \( z \) has been extracted through finite-size scaling of the energy gap, but calculating excited states using DMRG is often challenging. In contrast, long-time quench dynamics with complex-time allows direct inference of \( z \) from dynamics.
Secondly, dynamical structure factors $S(k,\omega)$ of the XXZ model under the periodic boundary condition are computed both in gapped and gapless phases. Specifically, in the XXZ model case, employing the complex-time evolution enables a direct computation of the spin excitations in the full momentum and frequency domain over the whole parameter space. We observe the transition from the gapped antiferromagnetic to the intermediate gapless phases, which eventually leads to the ferromagnetic polarized phase as the external field enhances. In both cases, the reduction of the entanglement entropy, which serves as a proxy for the required bond dimension and hence the computational cost, is found significant in the complex-time evolution. 
Our findings suggest that the complex-time evolution can be employed for cost-efficient numerical studies of the dynamics of various interacting spin systems.

\section{\label{sec:level1}Method}
\label{sec:method}

We follow the complex-time evolution scheme proposed by Ref.~\onlinecite{Grundner2024}. Specifically, we compute the following dynamical spin structure factor,
 
\begin{align}
&G\left(t+i\left(\tau + \delta \tau \right) \right) \nonumber\\
&\equiv \langle\psi_{0}| \hat{S}^{\alpha}\hat{S}^{\alpha}(t+i(\tau+\delta\tau))|\psi_{0}\rangle \nonumber\\
&= \left[ e^{-(\hat{H}-E_{0})\delta\tau} \hat{S}^{\alpha} | \psi_0 \rangle \right]^\dagger e^{i(\hat{H}-E_{0})(t+i\tau)}\hat{S}^{\alpha}|\psi_{0}\rangle. 
\label{eq:te2}
\end{align}

Here, $\vert \psi_0 \rangle$ is the ground state of the Hamiltonian obtained using DMRG. 

In Ref.~\onlinecite{Grundner2024}, several complex-time evolution schemes were proposed for obtaining real-time spectra from complex-time evolution results. Here we adopt the extrapolation method, by extrapolating data from multiple complex‑time paths along the parallel contour.
Note that various complex-time-evolved data can be obtained from a single ket state $e^{i(\hat{H}-E_{0})(t+i\tau)}\hat{S}^{\alpha}|\psi_{0}\rangle$ on a complex-time path $t+i\tau$ (with fixed $\tau$). Afterwards, inner products between the ket and bra states $\left[ e^{-(\hat{H}-E_{0})\delta\tau} \hat{S}^{\alpha} | \psi_0 \rangle \right]^\dagger$ with several chosen values of imaginary time $\delta\tau$ are taken to compute the correlation functions on different complex contours.

Subsequently, as introduced in Ref.~\onlinecite{Grundner2024}, the correlation functions in the complex-time domain can be represented as a series expansion in terms of the imaginary time as follows, 
\begin{equation}
\begin{pmatrix}
    G(t+i\tau_{1}) \\
    G(t+i\tau_{2}) \\ 
    \vdots \\ 
    G(t+i\tau_{k})    
\end{pmatrix}
=
{\bf M}
\begin{pmatrix}
    G(t) \\
    h_{1}(t) \\
    \vdots \\
    h_{k-1}(t)
\end{pmatrix},
\end{equation}
where ${\bf M}$ is 
\begin{equation}
{\bf M} = 
\begin{pmatrix}
	1 & \tau_{1}/\tau_{\rm max} & (\tau_{1}/\tau_{\rm max})^2 & \dots & (\tau_{1}/\tau_{\rm max})^{k-1} \\
	1 & \tau_{2}/\tau_{\rm max} & (\tau_{2}/\tau_{\rm max})^2 & \dots & (\tau_{2}/\tau_{\rm max})^{k-1} \\
	\vdots & \vdots & \vdots & \ddots &\vdots  \\
	1 & \tau_{k}/\tau_{\rm max} & (\tau_{k}/\tau_{\rm max})^2 & \dots & (\tau_{k}/\tau_{\rm max})^{k-1}
\end{pmatrix}. 
\end{equation}
Note that $\tau_{\rm max}$ is the maximal $\tau_k$, and that coefficients $h_i(t)$ of the series expansion don't need evaluation. In addition, we chose the convention that $\tau_1 < \tau_2 < \cdots < \tau_k$ and that $\delta\tau \equiv \tau_{i+1} - \tau_i$ are the same for all $1 \leq i \leq k-1$. 

The real-time dynamical structure factor can be computed by inverting the above equation, 

\begin{equation}
	G(t) = \sum_{k=1}^{n}
	\left({\bf M}^{-1} \right)_{1k} G(t + i\tau_{k}).				
\label{eq3}
\end{equation}

For an accurate extrapolation of $G(t)$, a sufficient number of dynamical structure factor along different complex-time paths $G(t+i\tau_k)$ is necessary. In addition, oversized values of $\tau$ may hinder the accuracy due to the excessive suppression of excited states. On the other hand, too small values of $\tau$ may increase the computational cost to obtain the initial ket state upon which imaginary-time-evolved bra states are contracted. In this study, we computed 13 complex‑time‑evolved data sets, where $\tau_{min}$ and $\tau_{max}$ were chosen so that $G(i\tau_{\rm min})$ and $G(i\tau_{\rm max})$ produce 80\% and 50\% of the value of $G(0)$, respectively. 
To obtain stable extrapolation results, we employed a combinatorial averaging technique as presented in Ref.~\onlinecite{Grundner2024}. In this approach, the extrapolation procedure [Eq.~(\ref{eq3})] is applied to all possible combinations of $n$ data points selected from the full dataset (containing 13 data in our case), and the final result is obtained by averaging the resulting extrapolated data. $n = 6$ was chosen for this study. From these choices, sufficiently accurate extrapolations onto the real-time axis could be obtained.

For this study, we used the Julia-based {\sc itensor} library \cite{itensor}. For ground-state DMRG calculations, we allowed the bond dimension to increase up to $m = 1000$ and the truncation threshold was kept as $w = 10^{-10}$. During the time-evolution, truncation threshold of $w_{t} =10^{-8}$ was employed, except for one of the XXZ calculations with $h = 2.0$ where $w_{t} = 6\times10^{-8}$ was used for computational issues.  
%
%
The 2-site TDVP method as implemented in {\sc itensor} was employed for real- and complex-time evolutions\cite{Yang2020}. Note that, linear prediction methods can be combined with the complex-time evolution to achieve better accuracy in the low-frequency regime.

\begin{figure}
\includegraphics[width=0.5\textwidth]{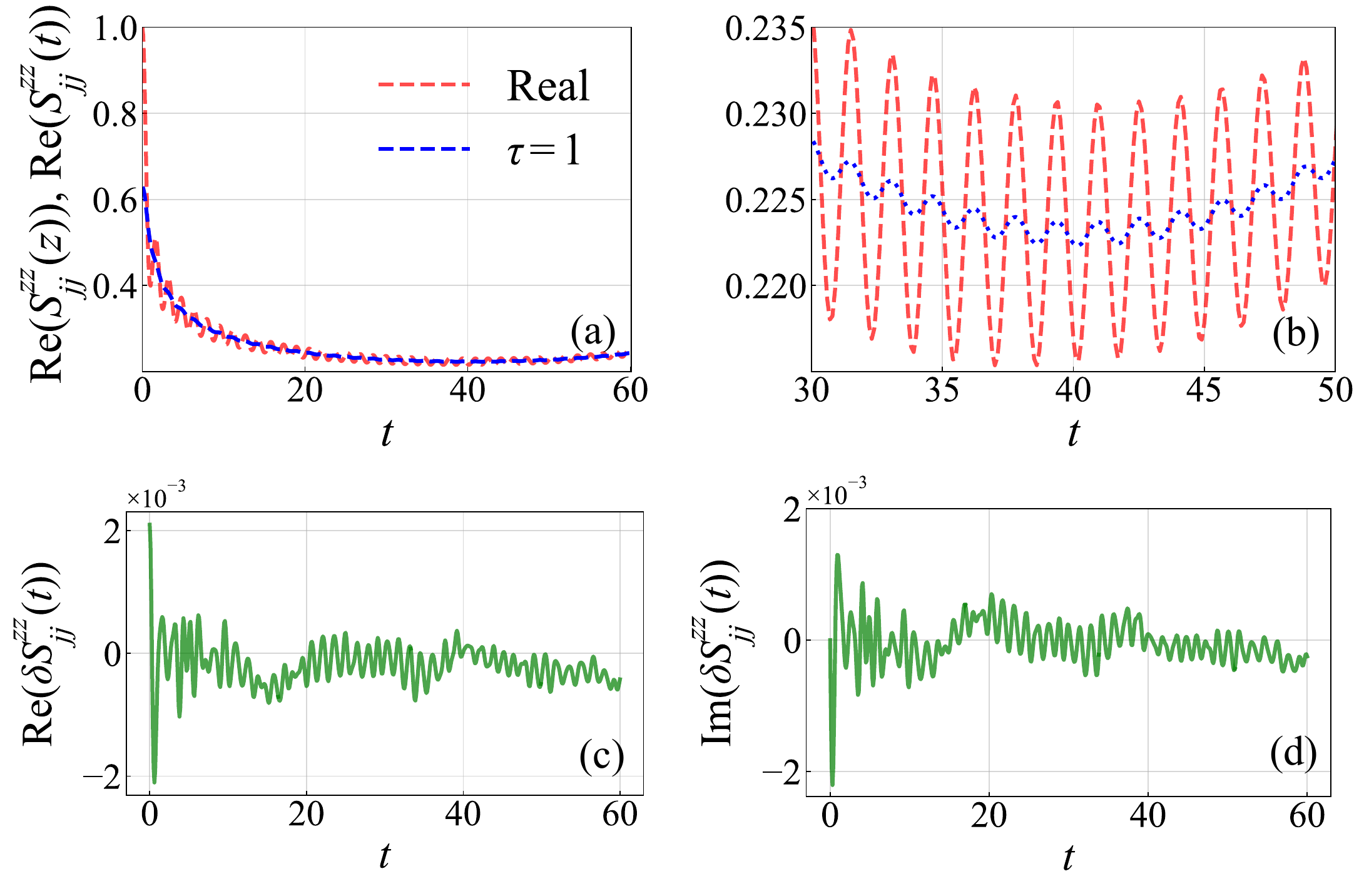}
\caption{(Color online) (a,b) ${\rm Re} (S_{jj}^{zz} (z))$ for each complex-time contour $z \equiv t+i\tau$, where (b) shows a magnified view of the interval $30 < t < 50$.
Panels (c) and (d) shows ${\rm Re} \delta S^{zz}_{jj}(t)$, and ${\rm Im} \delta S^{zz}_{jj}(t)$, respectively, where $\delta S^{zz}_{jj}(t) \equiv  S_{jj}^{zz} (t) -  S_{jj}^{zz} (\bar{t})$. Here $S_{jj}^{zz} (\bar{t})$ denote real-time-extrapolated data from the complex-time-evolved ones. 
}

\label{fig1}
\end{figure}

\section{\label{sec:level2}Transverse Field Ising Model}

First, we present the results on the one-dimensional (1D) $S=1/2$ TFIM. The Hamiltonian is given by 
\begin{equation}
	H = -J\sum_{i}S_{i}^{z}
	S_{i+1}^{z} - g\sum_{j}S_{j}^{x},
\end{equation}
where $S_i^\alpha$ ($\alpha=x,y,z$) is the $\alpha$-component of the spin-1/2 operator at site $i$. It exhibits a phase transition at \( J = 2g \), a critical point characterized by the $(1+1)D$ Ising universality class\cite{Sachdev_1999, park2015, Pang2019}. In this work, we chose $J=4$ and $g = 2$. 
To eliminate the edge effect, we adopted the periodic boundary condition (PBC). The following spin autocorrelation function was computed;
\begin{equation}
	S^{zz}_{jj}(t) = \langle\psi_{0}|S^{z}_{j}(t)S^{z}_{j}(0)|\psi_{0}\rangle,
\end{equation}
where \( |\psi_0\rangle \) is the ground state obtained by DMRG, and we set \( \hbar/2 = 1 \) for simplicity. Although \( j \) denotes the site index, the autocorrelation function is independent of \( j \) because the PBC is applied.
The time step was chosen as $t_{\rm step} = 0.1$ and the duration of the time evolution $t_{\rm max}$ was set to minimize the finite size effect.

Figure~\ref{fig1}(a) and (b) present the real part of $S^{zz}_{jj}(t)$ and $S^{zz}_{jj}(t+i\tau)$ before the finite-size scaling ($\tau=1$), where Fig.~\ref{fig1}(b) shows a magnified view of the interval $30 < t <50$. Note that the oscillatory behavior is reduced in the complex-time-evolved autocorrelation $S^{zz}_{jj}(t+i\tau)$, since the imaginary component $\tau$ in the time evolution effectively suppresses the weights of excited states. Figure~\ref{fig1}(c) and (d) show the real and imaginary components of the difference between $S^{zz}_{jj}(t)$ and $S^{zz}_{jj}(\bar{t})$, respectively, where $S^{zz}_{jj}(\bar{t})$ is obtained from the extrapolation of $S^{zz}_{jj}(z)$ to the real axis as described in Sec.~\ref{sec:method}. The results demonstrate that this method yields sufficiently accurate real-time data.

\begin{figure}
\includegraphics[width=0.5\textwidth]{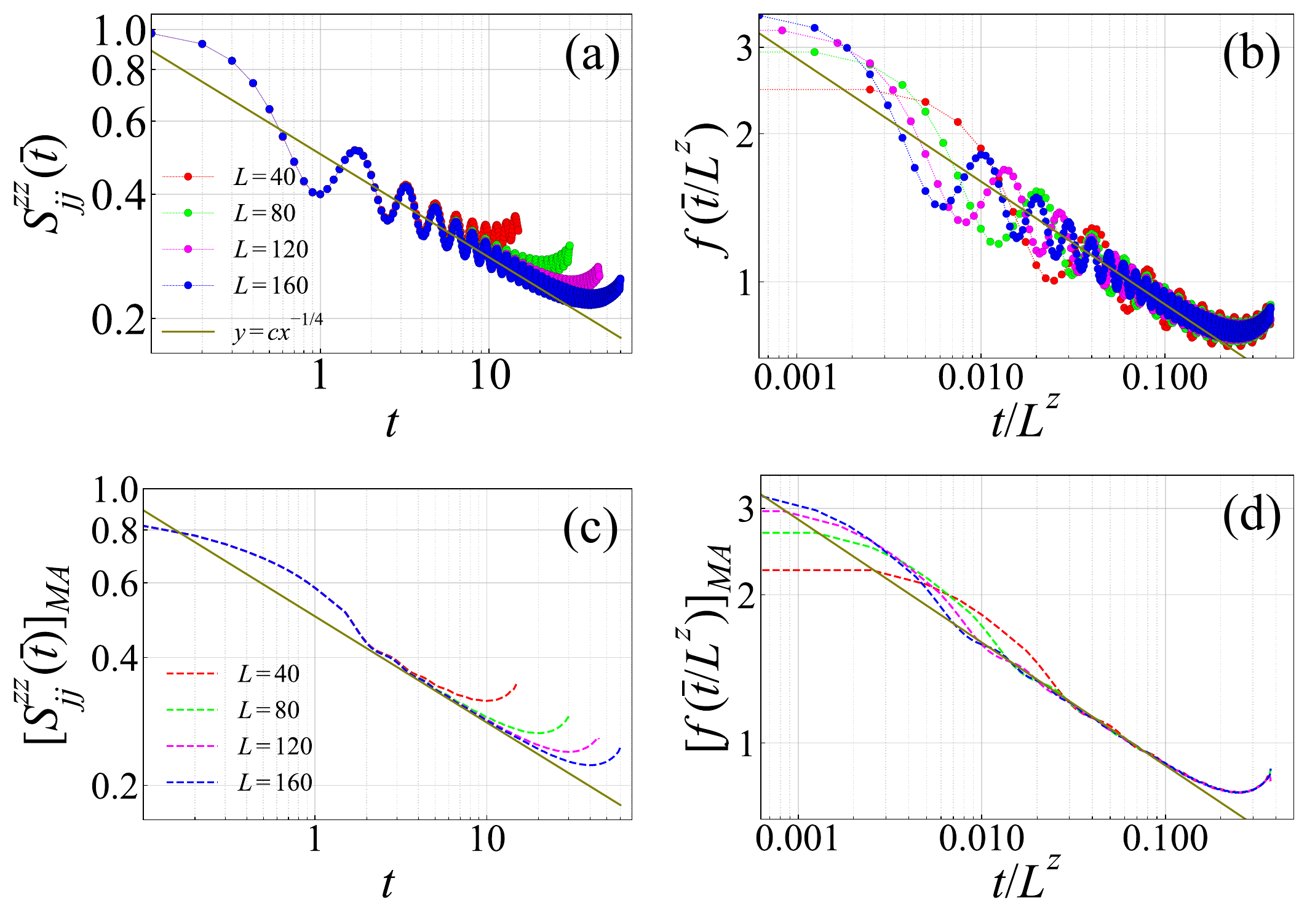}
	\caption{(Color online) (a) and (b) represent log-log plots of ${\rm Re}(S^{zz}_{jj}(t))$ before and after the finite-size scaling, respectively. (c) and (d) are after applying the moving average. (The constant $c$ was set to $0.5$ in panels (a) and (c), and to $288$ in panels (b) and (d).)
    }
\label{fig2}
\end{figure}
\begin{figure}
\includegraphics[width=0.48\textwidth]{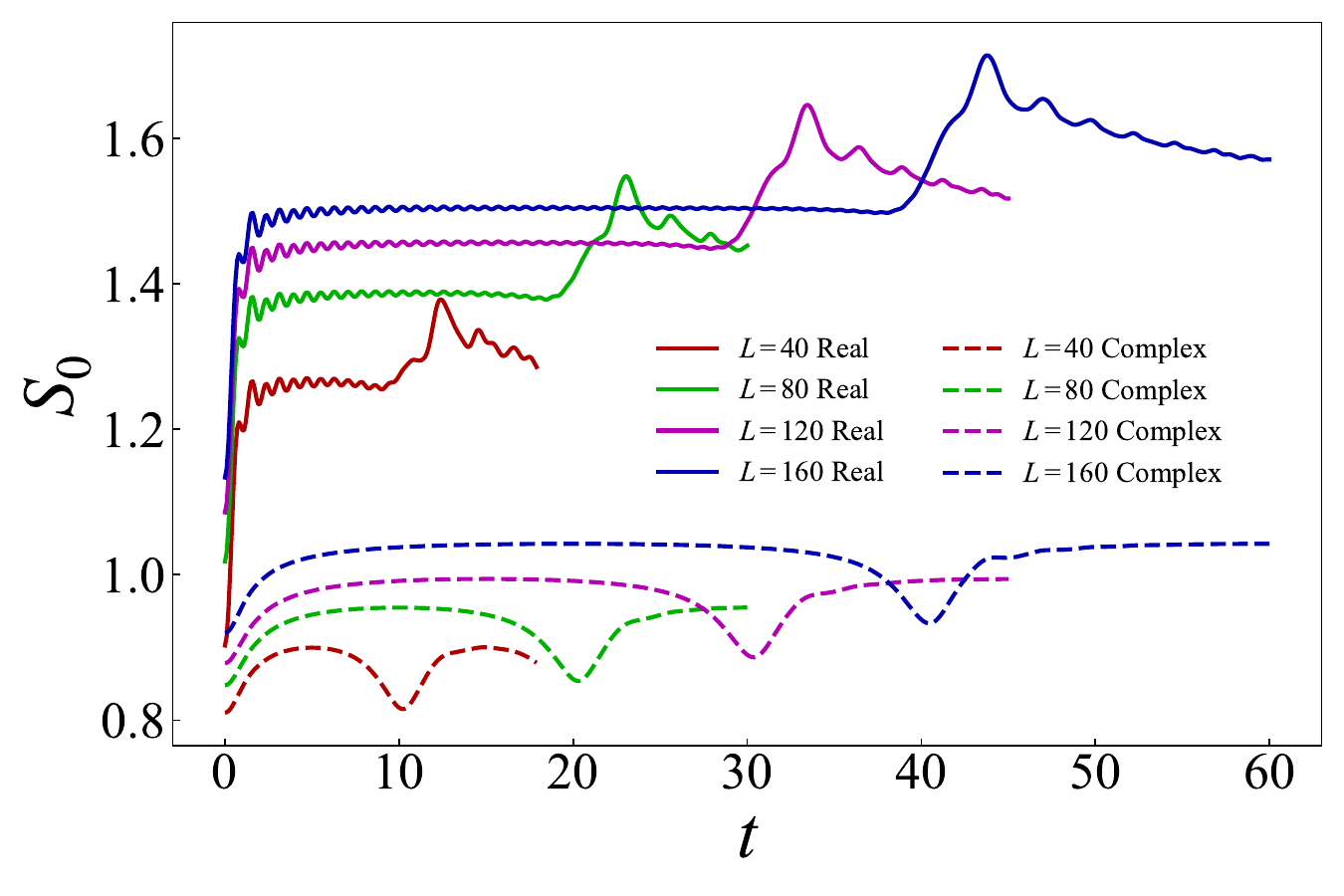}
\caption{(Color online) Comparison of bipartite entanglement entropy between real-time evolution (solid lines) and complex-time evolution (solid lines, all with $\tau=1$) for different system sizes.}
\label{fig4}
\end{figure}

For the estimation of the dynamical critical exponent, applying finite-size scaling is necessary.
At the quantum critical point, the behavior of this correlation function is governed by the scaling properties of the order parameter, which serves as the primary order parameter in the model. From the Ising conformal field theory, the order parameter  has a known scaling dimension  $ \Delta = 1/8$.
This scaling dimension determines the long-time decay of the correlation function in an infinite system. Specifically, the correlation function follows a power-law decay given by:
\begin{align}
    S^{zz}_{jj}(t) \sim t^{-2\Delta}.
\end{align}
To generalize this behavior to a finite-size system of length \( L \), we invoke the finite-size scaling theory. The autocorrelation function in a system of finite size obeys a scaling form:
\begin{align}
\label{eq:scaling}
S^{zz}_{jj}(t, L) = L^{-b} f\left(\frac{t}{L^z}\right),
\end{align}
where \( z \) is the dynamical critical exponent that relates time and length scales at criticality, \( b = 2\Delta z \) is a scaling exponent associated with the order parameter, and \( f(x) \) denotes a universal scaling function that depends only on the ratio \( x = t/L^z \).

Figure~\ref{fig2}(a) shows the log-log plot of the extrapolated autocorrelation function $S^{zz}_{jj}(\bar{t}
)$  from the complex-time evolution with respect to varying system size at the critical point. The finite-size-scaled ones following the universal scaling relation Eq.~\eqref{eq:scaling} are plotted in Fig.~\ref{fig2}(b).  
The fast oscillations in Fig.~\ref{fig2}(a) and (b), where the frequencies are $L$-independent and determined by the size of the transverse field $g$, are identified as harmonic-like amplitude modes away from the field-induced energy minima. Since they are not relevant to the dynamic critical behavior of the system, we smoothen them out by applying a moving average filter in the real-time domain. The results are shown in Fig. \ref{fig2}(c) and (d) (the former and latter corresponding to before and after applying the finite-size scaling, respectively). Note that the upturn of the $S^{zz}_{jj}(\bar{t})$ is a finite-size effect, and after the finite-size scaling all data with different values of $L$ collapse into a single curve.
A clear power-law behavior can be observed from Fig. \ref{fig2}(d), and we find that $\eta/z\approx 0.24$, which is in good agreement with the established theoretical values $z=1$ and $\eta = 0.25$.

Lastly, to estimate the reduction of the computational cost after applying the complex-time evolution, we calculated the change of bipartite entanglement entropy $S_0$ as a function of time. Figure~\ref{fig4} depicts these results, which shows that the growth of the entanglement entropy is suppressed in complex-time evolution cases. It is also checked that the decrease of the entanglement entropy in the complex-time evolution arises from the suppression of the singular values and the resulting reduction of the bond dimension, which leads to cheaper computational costs [see Sec.~\ref{sec:app} for further details]. This nicely presents the efficiency of the complex-time evolution methods in studying critical systems.

\begin{figure}
\includegraphics[width=0.5\textwidth]{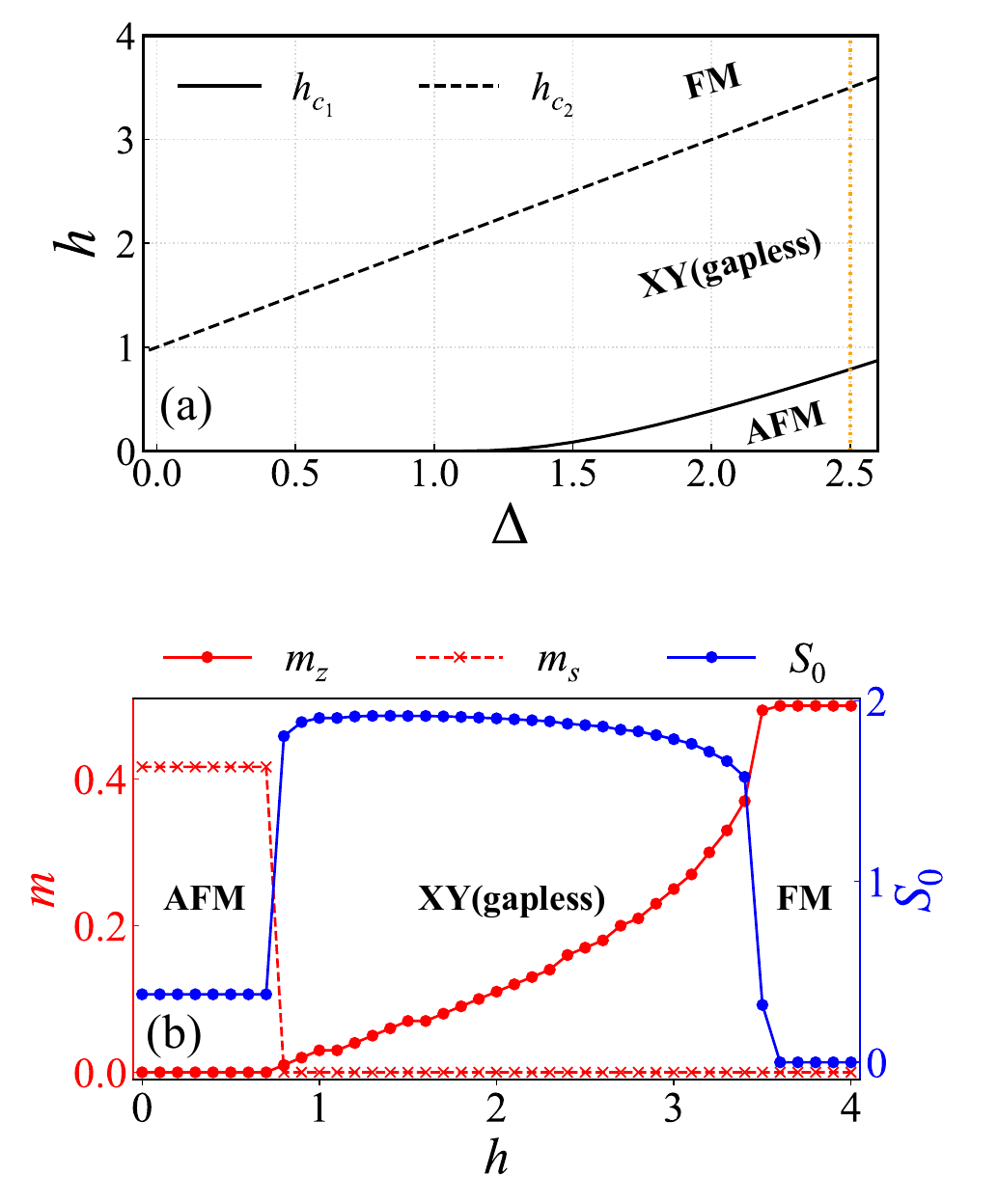}
\caption{(Color online) (a) Phase diagram of the one-dimensional XXZ model in the $\Delta-h$ space, where phase boundaries between three phases, Eq.~(\ref{eq:hc1}) and (\ref{eq:hc2}), are depicted as solid and dashed lines respectively.
(b) shows both the z-direction magnetization $m_z$ and the staggered magnetization $m_s = 1/L\sum_{i}(-1)^{i}S^z_{i}$, each serving as an order parameter as a function of $h$ for a fixed $\Delta$. As in (a), the solid and dashed lines indicate the critical points.}
\label{phase_diagram}
\end{figure}

\section{\label{sec:level3}XXZ Model}

For the second benchmark system we chose the one-dimensional $S=1/2$ XXZ model, where a XY gapless phase separates the antiferromagnetic (AFM) and the fully-polarized ferromagnetic (FM) phases in the presence of external Zeeman field [see Figure\,\ref{phase_diagram}\,(a)]. 
While the gapped state of this model has been thoroughly studied in various researches, the dynamics of the gapless state has been under-explored\cite{mahdavifar2007scaling,Bruognolo2016,Franchini2016,Rakov2015}.

The Hamiltonian of the XXZ model is as follows:
\begin{equation}
	H = {J\sum_{i}(S^{x}_{i}S^{x}_{i+1}+S^{y}_{i}S^{y}_{i+1})+\Delta S^{z}_{i}S^{z}_{i+1} -h\sum_{j}S^{z}_{j}}.
\end{equation}
The corresponding phase diagram with $J=1$ is shown in Fig.\,\ref{phase_diagram}\,(a). For $\Delta > 1$, this model has two critical points at \cite{yang1966}
\begin{equation}
	h_{c_{1}} = {\pi \sinh{(\arccos{\Delta})}\over\arccos{\Delta} \sum_{n=-\infty}^{\infty}{\rm sech}\left\{{\pi^{2}\over 2 \arccos\Delta}(1+2n)\right\}}
    \label{eq:hc1}
\end{equation}
and
\begin{equation}
	h_{c_{2}} = (1+\Delta)
    \label{eq:hc2}
\end{equation}
Here we chose $\Delta=2.5$ and studied spin dynamics of the AFM, gapless XY, and the FM phases. The critical fields for the transitions from AFM to XY and from XY to FM phases are $h_{c_{1}}\approx0.789$ and $h_{c_{2}}=3.5$, respectively. Fig.\,\ref{phase_diagram}\,(b) depicts the evolution of the spin ordering as a function of $h$ at $\Delta$ = 2.5, which shows three phases of the model including the gapless phase in the region between $h_{c_{1}}$ and $h_{c_{2}}$. Note that the gapless phase is strongly disturbed by applying the OBC due to the breaking of the translational symmetry at both ends of the chain and the following occurrence of edge modes. 

\begin{figure}
\includegraphics[width=0.5\textwidth]{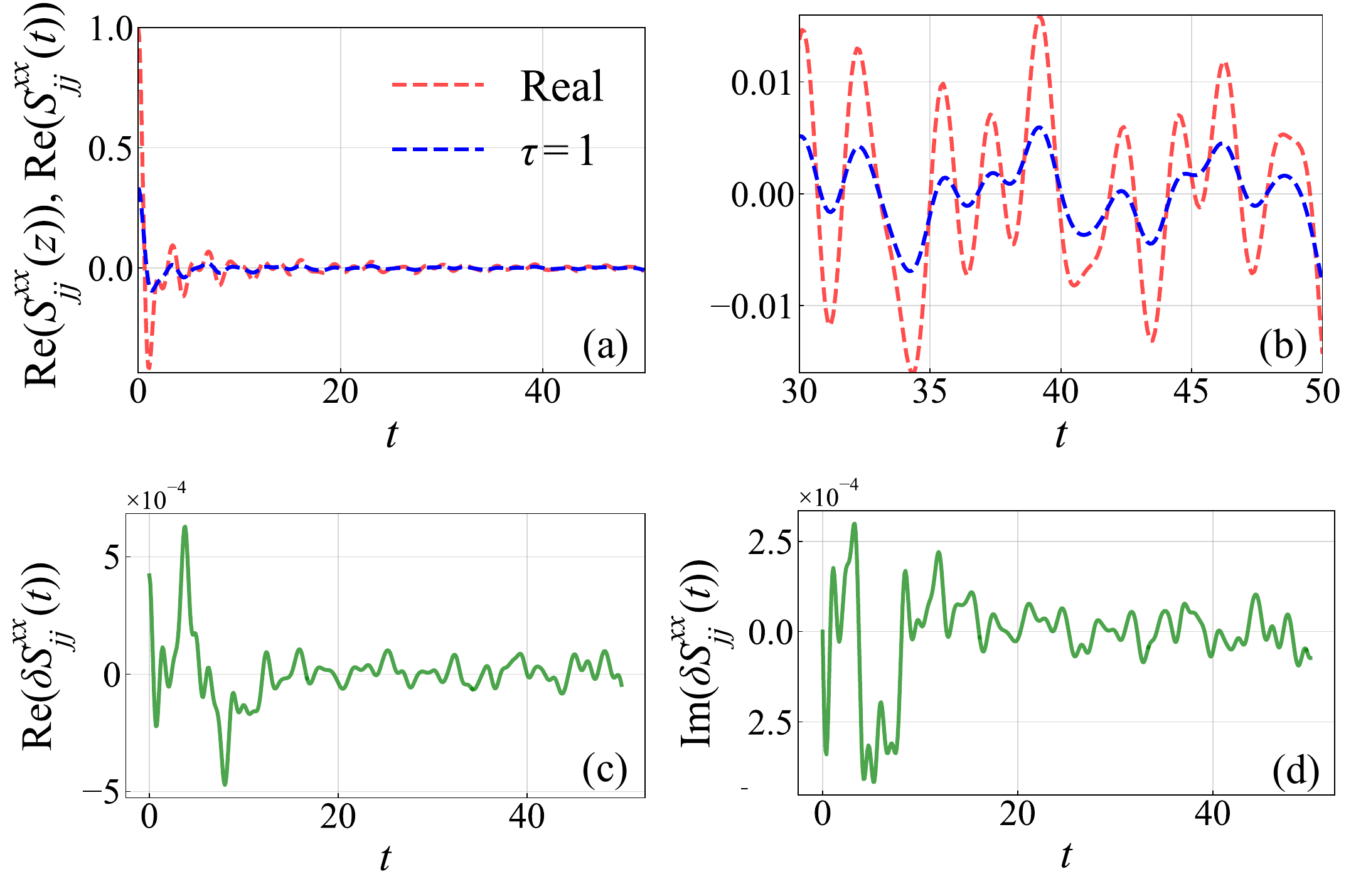}
	\caption{
    (Color online) (a,b) ${\rm Re} S_{jj}^{xx} (z)$ for each complex-time contour $z \equiv t+i\tau$ at $h = 0.8$, where (b) shows a magnified view of the interval $30 < t < 50$. Panels (c) and (d) shows ${\rm Re} \delta S^{zz}_{jj}(t)$, and ${\rm Im} \delta S^{zz}_{jj}(t)$, respectively.}

    \label{autocorrxxz}
\end{figure}

The spin autocorrelation function for the $x$-component of the spin can be expressed as
\begin{align}
	S^{xx}_{jj'}(t) = \langle\psi_{0}|S^{x}_{j}(t)S^{x}_{j'}(0)|\psi_{0}\rangle
    \label{eq:xcor}
\end{align}
Below we examine the local and momentum-resolved spectral functions; 

\begin{align}
	S^{xx}_{jj}(\omega) &= \int^{\infty}_{-\infty}dt e^{i\omega t}e^{-\epsilon(t)} S_{jj}^{xx}(t) \nonumber\\
	S^{xx}(k,\omega) &= \sum_{j}e^{-ika_{j}}\int^{\infty}_{-\infty}dt e^{i\omega t}e^{-\epsilon(t)} S_{0j}^{xx}(t),
\end{align}
where $e^{-\varepsilon(t)}$ is a Gaussian filter function, defined by $\varepsilon(t) = \alpha \times (t/t_{\rm max})^2$ where $\alpha=2$ in our calculations, and $a_j$ is the distance between the site 0 and $j$. 
As mentioned earlier, time evolution in the gapless state is limited by a large entanglement, making it efficient to obtain real-time data through complex-time evolution between $h_{c_{1}}$ and $h_{c_{2}}$. Note that we set $t_{\rm step} = 0.1 $ and $t_{\rm max}= 50$ keeping the maximum bond dimension $m_{\rm max} = 1000$.

\begin{figure}
\includegraphics[width=0.5\textwidth]{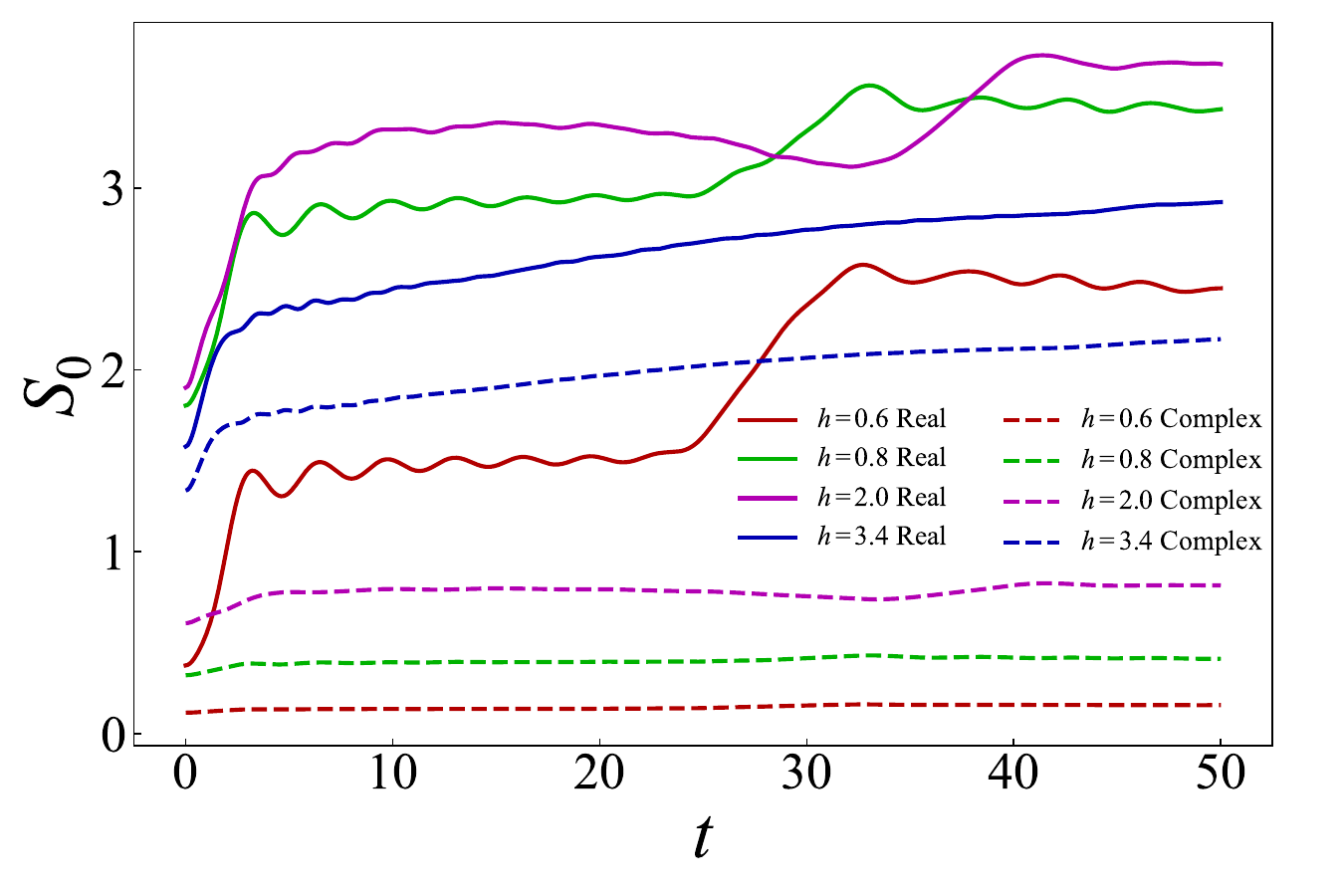}
	\caption{
    (Color online) Comparison of bipartite entanglement entropy between real-time evolution (solid lines) and complex-time evolution (dashed lines, all with $\tau=1$) for different strength of $h$ in the XXZ model case.}
    \label{entanglexxz}
\end{figure}

\begin{figure*}
\includegraphics[width=0.9\textwidth]{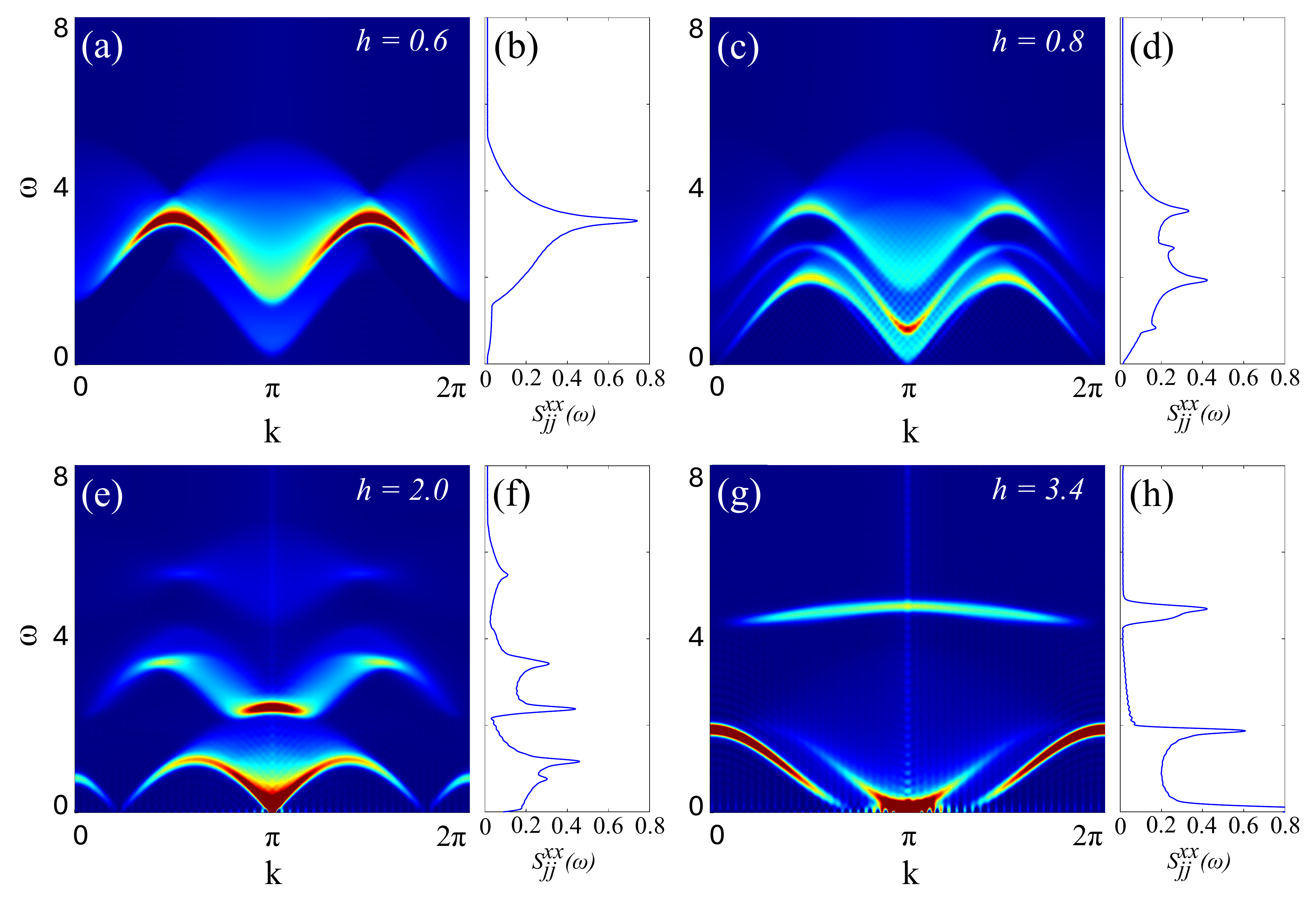}
	\caption{(Color online) (a-h) Dynamical structure factors $S^{xx}(k,\omega)$ and their local spectral functions $S^{xx}(\omega)$ at different field strengths. (a) and (b) represent the gapped spectrum before the phase transition, while (c)-(h) show spectra within the gapless phase.}
\label{Spectral}
\end{figure*}

Figure \ref{autocorrxxz} compares the real-time- and complex-time-evolved autocorrelation functions. This indicates that the extrapolation yields highly accurate results, even in the presence of intricate oscillatory behavior, suggesting an improvement over the widely used linear prediction (LP) method for studies of dynamics\cite{Tian2021}. Furthermore, Figure~\ref{entanglexxz} shows that the entanglement entropy remains remarkably low and does not grow during complex-time dynamics in the both the gapped and gapless phases. Note that at $h=3.4$, the reduced efficiency of complex-time evolution is currently hypothesized to stem from a large number of Goldstone modes in the system (see Fig.~\ref{Spectral}(g) and (h)); however, further investigation is necessary.

From Fig.~\ref{Spectral}, direct observations of the gapped state prior to the phase transition and the subsequent gapless state can be made. 
Initially, Fig.~\ref{Spectral}(a) and (b) shows the magnon gap and the two-spinon continuum in the AFM phase. 
Subsequently, Fig. \ref{Spectral}(c) and (d) depict the state immediately following the phase transition induced by $h$, characterized as an XY phase and confirmed to be gapless. Near the critical point, the emergence of multiple-spinon excitations, in addition to the two-spinon excitation, becomes prominent. This results in multiple continua separated by approximately $\Delta$, as explicitly illustrated in Fig.~\ref{Spectral}\,(c).
Employing the Jordan-Wigner transformation analogy, this transition indirectly reveals a filling-controlled Mott insulator-to-metal transition in the presence of the next-nearest-neighbor Coulomb repulsion, because increasing $h$ corresponds to enhancing the chemical potential in the fermion analogue. The observation of multiple spectra along the $\omega$-axis arises because the $x$-direction correlation function, viewed from the spinless fermion perspective, captures both the (quasi-)particle and hole excitations projected onto the positive frequency domain. Based on this interpretation, the spectrum with the highest frequency can be identified as the lower Hubbard band, whereas the middle and lowest ones correspond to the upper Hubbard band and the coherent metallic band, respectively.

In the XY phase, the dynamical spin structure factor is characterized by a gapless spinon continuum and exhibits singular behavior near its lower boundary, a feature clearly captured by our calculations\,[see Fig.\,\ref{Spectral}\,(e)]. Furthermore, around $\omega\approx h$ in Fig.\,\ref{Spectral}\,(d), we observe a pronounced peak, which may indicate the formation of a magnon bound state.
In the subsequent panels, Fig.\,\ref{Spectral}\,(e) through (h), the system evolves from the XY-like phase to spin-polarized one. Notable, at $h = 3.4$ [Fig.~\ref{Spectral}(g) and (h)], a sharp increase in the Goldstone mode is observed. As the spins become nearly fully aligned, the spectral profile begins to exhibit a quadratic shape, a characteristic feature of the ferromagnetic (FM) phase. the resulting proliferation of low-energy modes may be the source of the computational cost increase in our complex-time evolution at $h=3.4$ [see the blue dashed curve in Fig.~\ref{entanglexxz}].

The equivalence between the spin model and the spinless fermion model of the XXZ model, as elucidated through the Jordan-Wigner transformation, has been recognized for some time. However, by applying complex-time evolution starting from the spin system, it becomes possible to directly observe behavior analogous to the Mott insulator–metal transition seen in fermionic systems. Furthermore, when considering entanglement entropy, the significance lies in the enhanced efficiency of this approach for verifying such physical quantities compared to the commonly employed real-time evolution method.

\section{\label{sec:level4}Discussion and conclusion}

In this study, we have demonstrated the effectiveness of using complex-time evolution to study the dynamics of one-dimensional spin models, particularly the TFIM and the XXZ model.	
Using the parallel contour method, we effectively suppressed the increase in entanglement entropy, a major challenge in real-time evolution studies.	
The results confirm that complex-time evolution, combined with the extrapolation method, offers a robust framework for examining the dynamical properties of systems with large entanglement entropy.	 
We successfully re-evaluated the dynamic critical exponent $z$ for the TFIM, and investigated the dynamical structure factor $S^{xx}(k,\omega)$ near the critical point in both gapped and gapless states of the XXZ model.
   The findings suggest that this approach not only enhances computational efficiency but also maintains high accuracy, making it a valuable tool for further studies on the dynamics of complex quantum systems. The suppression of entanglement entropy growth observed in the XXZ model compared to the TFIM case further highlights the potential of complex-time evolution in handling systems with significant entanglement.  Moreover, the methodologies presented here are not limited to the AIM model but can be extended to other models with significant entanglement entropy. This paves the way for more 
comprehensive and efficient investigations into the dynamic behaviors of various quantum systems.

\begin{acknowledgments}
We thank Jan von Delft, Wei-Lin Tu, Edwin M. Stoudenmire, Seung-Sup B. Lee, Aaram J. Kim, Young-Woo Son, Hosub Jin, Choong H. Kim, Ara Go, and Sangkook Choi for fruitful discussions. This work was supported by the Basic Science Research Program through the National Research Foundation of Korea funded by the Ministry of Science and ICT [Grant No. NRF-2020R1C1C1005900, RS-2023-00220471]. 
HSK additionally appreciates the hospitality of the Center for Theoretical Physics of Complex Systems at the Institute of Basic Science, Korea, and of the Asia Pacific Center for Theoretical Physics, Korea, during the completion of this work [APCTP-2025-C01].
\end{acknowledgments}

\begin{appendix}

\section{\label{sec:app}Singular values and bond dimension in complex-time evolution of the TFIM}

\begin{figure}
\includegraphics[width=0.45\textwidth]{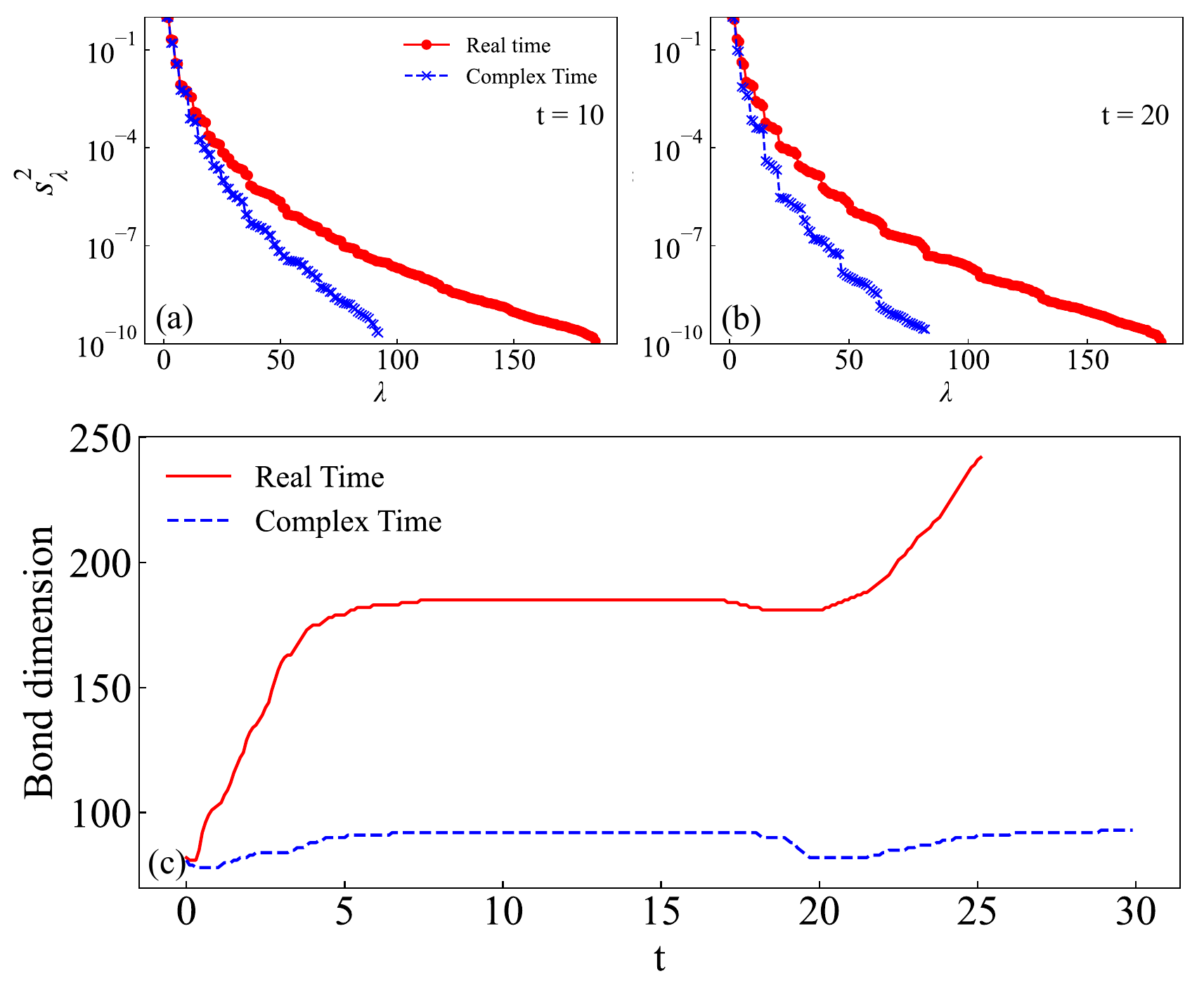}
	\caption{(Color online) Panel (a) and (b) plot the squared bipartite singular-value spectra of the time-evolved ground state of the TFIM model at $t = 10$ and $t = 20$, respectively (y-axis being in a logarithmic scale). (c) illustrates the evolution of the maximum bond dimension as a function of time.}
    \label{Appfig1}
\end{figure} 

In general, when using a fixed truncation threshold, a reduction in the entanglement entropy typically leads to a corresponding decrease in the bond dimension. However, if the singular values decay more slowly, a reduction in entanglement entropy under the same truncation threshold may not necessarily result in a similar decrease in bond dimension. To examine this aspect, we compare the distribution of singular values in matrix product states evolved via complex-time and real-time evolution.

Figure~\ref{Appfig1} presents the singular-value spectra and the corresponding bond dimensions obtained from both real-time and complex-time evolutions, viewed from this perspective. The simulations were performed at the critical point of the transverse-field Ising model (TFIM) for a chain of length $N = 80$, with the truncation error maintained at $w_t = 10^{-10}$ throughout the time evolution.

As shown in Figs.~\ref{Appfig1}(a) and (b), which display the squared singular-value distributions at $t=10$ and 20, respectively, the singular values from the complex-time evolution exhibit much steeper decline compared to those from real-time evolution. Consequently, the time-dependent bond dimension in Fig.~\ref{Appfig1}(c) closely follows the behavior of the entanglement entropy shown earlier in Fig.~\ref{fig4}, thereby reaffirming the advantage of the complex-time evolution scheme.

\end{appendix}

\bibliography{apssamp}
\end{document}